\documentclass[12pt]{article}
\usepackage{times}
\usepackage{geometry}
\usepackage[utf8]{inputenc}
\usepackage{enumitem,amssymb}
\usepackage{ragged2e}
\newlist{thematic}{itemize}{8}
\setlist[thematic]{label=$\square$}
\usepackage{pifont}
\usepackage{natbib}
\usepackage{graphicx}

\usepackage{rotating}

\begin{document}
\raggedright
\huge
Astro2020 Science White Paper \linebreak

Extragalactic Proper Motions:  Gravitational Waves and Cosmology \linebreak
\normalsize

\noindent \textbf{Thematic Areas:} \hspace*{60pt} $\square$ Planetary Systems \hspace*{10pt} $\square$ Star and Planet Formation \hspace*{20pt}\linebreak
$\square$ Formation and Evolution of Compact Objects \hspace*{31pt} \makebox[0pt][l]{$\square$}\raisebox{.15ex}{\hspace{0.1em}$\checkmark$} Cosmology and Fundamental Physics \linebreak
  $\square$  Stars and Stellar Evolution \hspace*{1pt} $\square$ Resolved Stellar Populations and their Environments \hspace*{40pt} \linebreak
  $\square$    Galaxy Evolution   \hspace*{45pt} $\square$             Multi-Messenger Astronomy and Astrophysics \hspace*{65pt} \linebreak
  
\textbf{Principal Author:}

Name:	Jeremy Darling\footnote{The authors acknowledge support from the NSF grant AST-1411605
and the NASA grant 14-ATP14-0086.}
 \linebreak						
Institution:  University of Colorado
 \linebreak
Email:  jeremy.darling@colorado.edu
 \linebreak
Phone:  303 492 4881
 \linebreak
 
\textbf{Co-authors:} Alexandra Truebenbach \&  Jennie Paine (University of Colorado)
  \linebreak 

\vspace{-10pt}
\begin{figure}[h]
  \includegraphics[width=1.0\textwidth,trim=90 110 400 260,clip]{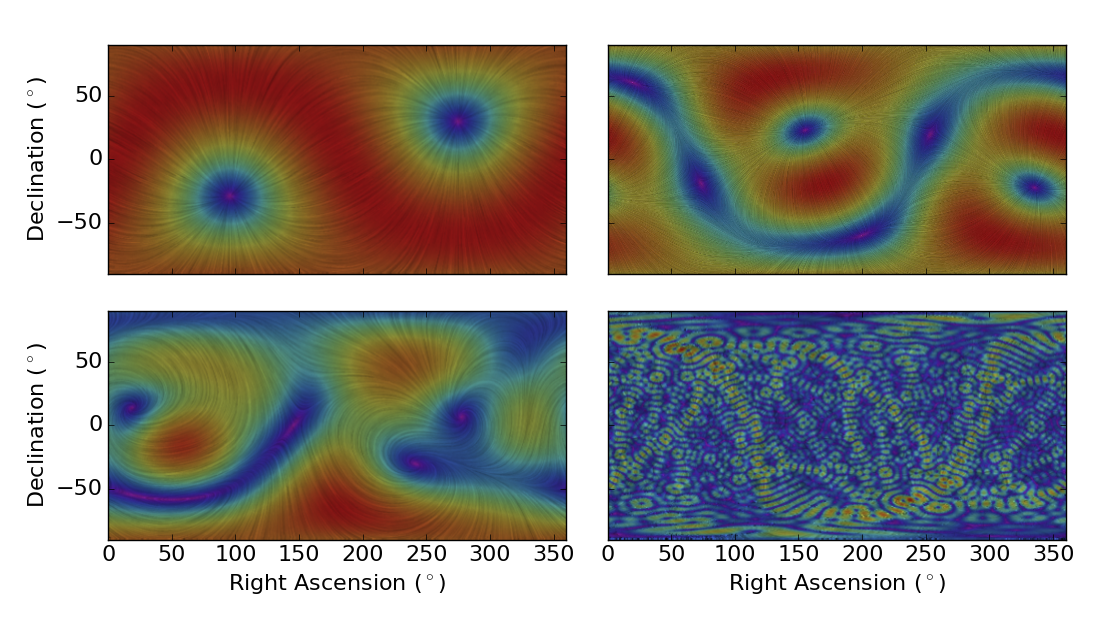}
\end{figure}
\vspace{-5pt}

\textbf{Abstract:}
Extragalactic proper motions can reveal a variety of cosmological and local phenomena over a range of angular scales.  These include 
observer-induced proper motions, such as the secular aberration drift caused by the solar acceleration about the Galactic Center and
 secular extragalactic parallax resulting from our motion with respect to the cosmic microwave background rest frame.  Cosmological 
effects include the isotropy of the Hubble expansion, transverse peculiar velocities induced by large scale structure, the real-time evolution of the 
baryon acoustic oscillation, and long-period gravitational waves that deflect light rays, producing an apparent quadrupolar proper motion
pattern on the sky.  We review these effects, their imprints on global correlated extragalactic proper motions, their expected amplitudes, the current best measurements
(if any), and predictions for {\it Gaia}.  Finally, we describe key ground- and space-based observational requirements to measure or constrain these proper motion signals down to amplitudes of
$\sim$0.1~$\mu$as~yr$^{-1}$, or $\sim$0.7\% of $H_0$.\footnote{Portions of this science white paper were 
adapted from \citet{darling2018c} with permission from the publisher.}

\pagebreak

\setcounter{page}{1}

\section{Background}

\vspace{-6pt} 
The universe is dynamic, as we know from the Hubble expansion, but astronomers treat extragalactic
objects as fixed in the sky with fixed redshifts.  If measured with enough precision, however, nothing is constant:  all objects
will change their redshifts and angular positions at a rate of order $H_0 \simeq 7\times10^{-11}$~yr$^{-1}$.  The secular redshift drift caused by a non-constant expansion \citep{sandage1962} is of order 0.3~cm~s$^{-1}$~yr$^{-1}$ at $z\simeq1$ and may be measured by optical telescopes
using the Ly$\alpha$ forest or by radio telescopes using H{\small I} 21 cm or molecular absorption lines \citep{loeb1998,darling2012}.  Proper motions of 
extragalactic objects are caused by peculiar motions induced by large scale structure 
\citep{darling2013,darling2018b,truebenbach2018}, 
by primordial gravitational waves \citep[e.g.,][]{pyne1996,gwinn1997,book2011,darling2018}, 
by the recession of cosmic rulers such as the baryon acoustic oscillation, or by anisotropic expansion 
\citep{fontanini2009,quercellini2009,titov2009,darling2014}.

\hspace{10pt} 
Observer-induced proper motions are also possible:  these include the secular aberration drift caused by acceleration of the solar system barycenter about the Galactic Center \citep[e.g.,][]{bastian1995,eubanks1995}
and secular extragalactic parallax caused by motion with respect to the cosmic microwave background \citep[CMB;][]{ding2009}.

\hspace{10pt}  
Proper motions depict a discretely sampled vector field on the celestial sphere.  It 
is therefore natural to describe proper motions using vector spherical harmonics (VSH),
the vector extension of the scalar spherical harmonics used
to describe signals such as the CMB temperature pattern, the geoid, or equipotentials \citep{mignard2012}.  
VSH are characterized by their degree $\ell$ and 
order $m$, and resemble electromagnetic fields.  They can therefore be separated into curl-free (E-mode) and divergenceless (B-mode) 
vector fields that are typically connected to distinct physical phenomena.  The general method for characterizing a correlated proper motion 
field is to fit VSH to the observed proper motions and to calculate the power in each mode.

\vspace{-10pt}
\section{Expected (and Possible) Signals}\label{sec:signals}

\vspace{-6pt}  
Table \ref{Tab:Summary} summarizes the expected and possible global extragalactic proper motion signals.  This summary is likely to be 
incomplete.  Here we provide a brief description of each physical or observer-induced effect, and Section \ref{sec:reqs} discusses the key observational requirements for detecting or constraining these phenomena.

\vspace{-10pt}
\subsection{Secular Aberration Drift}

\begin{figure}[t!]
\includegraphics[width=1.0\textwidth,trim=20 30 20 50]{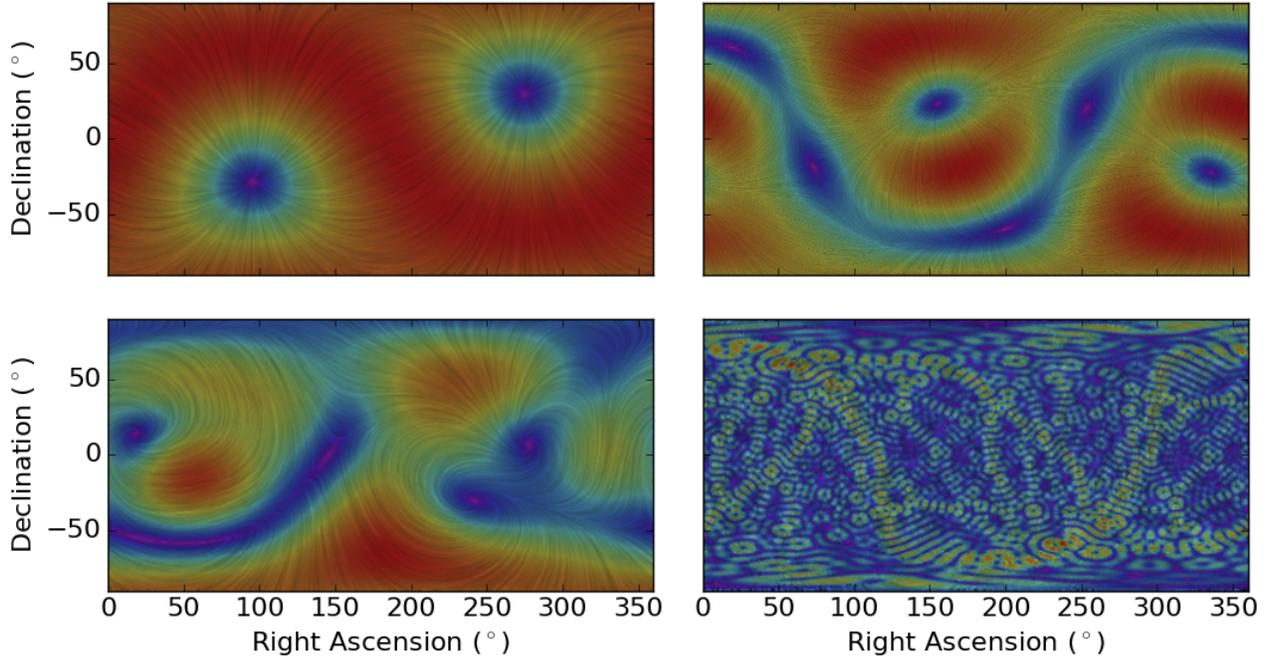}
\caption{
All-sky stream plots.  Streamlines indicate the vector field direction, and the colors indicate the vector amplitude, from 
violet (zero) to red (maximum).
Panels, clockwise from top left, indicate:
(1) the secular aberration drift dipole detected by \citet{truebenbach2017};
(2) a triaxial anisotropic Hubble expansion \citep[e.g.,][]{darling2014,paine2018};
(3) stochastic gravitational waves causing light deflection \citep[after][]{darling2018};
(4) BAO recession streamlines.}\label{fig:streamplots}
\vspace{-8pt}
\end{figure}

\vspace{-6pt} 
Aberration of light is caused by the finite speed of light and the motion of the observer with respect to a light source.  The resulting deflection 
of light scales as $\vec{v}/c$.  If the observer accelerates, then the aberration exhibits a secular drift, and objects appear to stream in the direction of the acceleration vector.  The solar system barycenter accelerates at roughly 0.7~cm~s$^{-1}$~yr$^{-1}$ as the Sun orbits the Galactic Center, resulting
in an apparent $\sim$5~$\mu$arcsec~yr$^{-1}$ E-mode dipole converging on the Galactic Center (Figure \ref{fig:streamplots}).  This has been detected in the proper motions of 
radio sources, first by \citet{titov2011}, using {\it a priori} knowledge of the expected effect, and recently without priors by 
\citet{truebenbach2017}.  
\citet{titov2018} have further developed the VLBI-specific methodology for extracting this signal.  

\vspace{-10pt}
\subsection{Secular Parallax}

\vspace{-6pt} 
The CMB shows a temperature dipole of 3.4 mK due to the motion of the solar system barycenter with respect to the CMB 
rest frame \citep[e.g.,][]{hinshaw2009}.  This amounts to a relative motion of $369\pm0.9$~km~s$^{-1}$ or 
78 AU~yr$^{-1}$.  This motion will 
induce a maximum secular parallax of 78~$\mu$arcsec~yr$^{-1}$~Mpc$^{-1}$:  nearby galaxies in directions perpendicular to the CMB
poles will show a reflex motion opposite our motion when compared to distant galaxies.  Precise proper motion measurements of galaxies
within $\sim$50~Mpc will allow a statistical detection of the distance-dependent E-mode dipole caused by secular parallax.  When 
referenced to this average dipole, it will be possible to detect the peculiar motions of individual galaxies as well as their geometric 
distances.  Secular parallax may provide a distance ladder-free method for measuring distances in the local universe (and thus $H_0$).

\vspace{-10pt}
\subsection{Rotation}

\vspace{-6pt}
As rotating observers, we use quasars to define a fixed, non-rotating reference frame in order to 
measure and monitor the rotation of the Earth \citep[e.g.,][]{mccarthy2004}.  We have no means to detect non-terrestrial rotation that has an axis close to polar, but
it is possible to detect or constrain rotation about axes that are not aligned with the polar axis.  While the question of a rotating universe 
can be nettlesome to contemplate and violates our assumption of cosmological isotropy, one can nonetheless make precise, 
sub-$\mu$arcsec~yr$^{-1}$ measurements of the effect because it would manifest as a B-mode dipole in the proper motion vector field.

\vspace{-10pt}
\subsection{Anisotropic Expansion}

\vspace{-6pt} 
In an isotropically expanding universe, objects move radially away from every observer (modulo small peculiar motions due to 
local density perturbations; see Section \ref{sec:lss}), so there will be no global correlated proper motions caused by 
isotropic Hubble expansion.  However, anisotropic expansion would cause objects to stream across the sky toward the directions of 
fastest expansion and away from directions of slowest expansion (Figure \ref{fig:streamplots}).  Assuming triaxial anisotropy, one can show that the 
resulting celestial proper motion pattern is completely described by an E-mode quadrupole \citep{darling2014}.  Fitting a
curl-free quadrupole to a proper motion field can therefore measure or constrain the (an)isotropy of the Hubble expansion without
{\it a priori} knowledge of $H_0$.  We can express the Hubble constant in terms of an angular rate, $H_0 \simeq 15$~$\mu$arcsec~yr$^{-1}$,
which means that a 10\% anisotropy would produce a quadrupole amplitude of 1.5~$\mu$arcsec~yr$^{-1}$.

\vspace{-10pt}
\subsection{Gravitational Waves}

\vspace{-6pt} 
Stochastic gravitational waves deflect light rays in a quadrupolar (and higher $\ell$) pattern with equal power in the E- and B-modes
\citep[Figure \ref{fig:streamplots};][]{pyne1996,gwinn1997,book2011}.  
Proper motion observations are sensitive to gravitational wave frequencies
$10^{-18}$~Hz~$< f < 10^{-8}$~Hz ($H_0$ to 0.3 yr$^{-1}$), which overlap the
pulsar timing and CMB polarization regimes, but uniquely span
about seven orders of magnitude between the two methods \citep{darling2018}.  The cosmic energy density of gravitational
waves $\Omega_{GW}$ can be related to the proper motion quadrupolar power $P_2$ or
variance $\langle\mu^2\rangle$ as
\begin{equation} 
  \Omega_{\rm GW} = {6\over5}\,{1\over4\pi}\,{P_2\over H_\circ^2} 
  = 0.00042\, {P_2\over (1\ \mu\rm{as\ yr}^{-1})^2}\,h_{70}^{-2}
  \sim \langle\mu^2\rangle/H_0^2
\end{equation}  
\citep{gwinn1997,book2011,darling2018}.  
Measuring or constraining the proper motion quadrupole power 
can therefore detect or place limits on primordial gravitational waves in a unique portion of the gravitational wave spectrum.

\vspace{-10pt}
\subsection{Large Scale Structure}\label{sec:lss}

\vspace{-6pt} 
The mass density distribution of large-scale structure reflects the mass power spectrum, the shape and evolution of which relies on cosmological parameters. The transverse peculiar motions of extragalactic objects can be used to measure the density perturbations from large-scale structure without a reliance on precise distance measurements. While line-of-sight velocity studies use distances to differentiate Hubble expansion from peculiar velocity, peculiar motions across the line-of-sight are separable from Hubble expansion because no proper motion will occur in a homogeneous expansion \citep{nusser2012,darling2013}.
Thus, one can employ a two-dimensional transverse velocity correlation function to connect proper motions
to large scale structure \citep{truebenbach2018}.  This same measure is straightforward to connect to the
theoretical matter power spectrum \citep{darling2018b,hall2018}.  Proper motions can  
thus test the shape of the mass power spectrum without a dependence on precise distance measurements or
reliance on a cosmological distance ladder.

\vspace{-10pt}
\subsection{Baryon Acoustic Oscillation Evolution}

\vspace{-6pt} 
The baryon acoustic oscillation (BAO) is a ``standard ruler'' arising from pre-CMB density fluctuations, which 
can be observed as an overdensity of galaxies on the scale of $\sim$150 Mpc, comoving \citep{eisenstein2005}.
At redshift $z=0.5$, the BAO scale subtends $\theta_{BAO} =  4.5^\circ$, which is equivalent to VSH degree $\ell\sim40$.  
Taking the time derivative, we obtain an expression for proper motion on these scales:
\begin{equation}
\mu_{BAO} = {\Delta\theta_{BAO}\over\Delta t_\circ} \simeq - \theta_{BAO}\ H_0 \simeq -1.2\ \mu{\rm as}\ {\rm yr}^{-1}.
\label{eqn:bao}
\end{equation}
The BAO evolution will therefore manifest as a convergent E-mode signal around $\ell\sim40$ (Figure \ref{fig:streamplots}).
To first order, the BAO scale depends on the expansion rate $H(z)$ and the angular diameter distance $D_A(z)$ at the observed
redshift, but the rate of change of this standard ruler is dominated by its recession (``receding objects appear to shrink'') and 
depends to first order on $H_0/D_A(z)$.  Detection of this effect relies critically on the sky density of sources, which must adequately sample
angular scales smaller than $4.5^\circ$.

\begin{sidewaystable}
\caption[Global Correlated Proper Motion Signal]{Global Correlated Proper Motion Signal\footnote{After \citet{darling2018c}.}}
\smallskip
\begin{center}
{\small
\begin{tabular}{lrccccccc}  
\hline
\noalign{\smallskip}
Effect & $\ell$ & Mode & Amplitude & Recent Measurement & Ref & {\it Gaia} & ngVLA or & Prev.\ Work \\
 & & & & & & & Space Astrometry & \\
  \noalign{\smallskip}
 & & & ($\mu$arcsec yr$^{-1}$)  & ($\mu$arcsec yr$^{-1}$)  & & (Predicted) & (Predicted)\\
\noalign{\smallskip}
\hline
\noalign{\smallskip}
Secular Aberration Drift & 1 &  E  &  $\sim$5  & $5.2\pm0.2$ & 1 & 10$\sigma$ & 50$\sigma$ & 2,3,4,5 \\
Secular Parallax                  & 1 & E  &   78~Mpc$^{-1}$  & ... & ... & $\sim$10$\sigma$ & $\sim$10$\sigma$~$^{(1)}$ & 6 \\
Rotation                             & 1 & B &    Unknown & $0.45\pm0.27$ & 7 & $<0.5$~$\mu$as yr$^{-1}$ & $<0.1$~$\mu$as yr$^{-1}$~$^{(2)}$ & 2,5,7 \\
Anisotropic Expansion       & 2  & E   &  Unknown & $<7$\% & 8 &  $<3$\% & $<0.7$\% & 8,9,10,11\\
Gravitational Waves           &  $\geq2$ & E+B & Unknown & $\Omega_{GW} < 6.4\times10^{-3}$ & 7 & $\Omega_{GW} < 4\times10^{-4}$& $\Omega_{GW} < 10^{-5}$ & 1,7,12,13,14,15,16 \\
Large Scale Structure          &      $\gtrsim 5$ & E & $-15$ to +5  & $8.3\pm14.9$ & 17 & 10$\sigma$ & $\sim$20$\sigma$~$^{(3)}$ & 17,18,19,20 \\
BAO Evolution                 & $\sim$40 & E  &  $-1.2$  at $z=0.5$  & ... & ... & 4$\sigma$ & $\sim$10$\sigma$~$^{(4)}$ & ... \\
\noalign{\smallskip}
\hline 
\noalign{\smallskip}
\end{tabular}
}
\end{center}
{\small
Notes:
1 -- Detection of the secular parallax is limited by the number of compact radio sources that can (and would) be 
monitored in the local volume.
2 -- Observations will only be sensitive to rotation axes that are not aligned with the Earth's rotation axis.
3 -- Detection of peculiar velocities associated with large scale structure will depend on the number of close galaxy pairs.
4 -- Detection of the BAO evolution will depend on the sky density of proper motion measurements.\\ 
References:
1 -- \citet{titov2018}
2 - \citet{titov2011};  
3 -- \citet{xu2012};
4 -- \citet{titov2013};
5 -- \citet{truebenbach2017};
6 -- \citet{ding2009};
7 -- \citet{darling2018}; 
8 -- \citet{darling2014};
9 -- \citet{chang2015};
10 -- \citet{bengaly2016}; 
11 -- \citet{paine2018};
12 -- \citet{braginsky1990};
13 -- \citet{pyne1996};
14 -- \citet{kaiser1997};
15 -- \citet{gwinn1997};
16 -- \citet{book2011};
17 -- \citet{darling2013};
18 -- \citet{darling2018b};
19 -- \citet{truebenbach2018};
20 -- \citet{hall2018}.}
\label{Tab:Summary}
\end{sidewaystable}

\vspace{-10pt}
\section{Key Observational Requirements}\label{sec:reqs}

\vspace{-6pt}
Both ground- and space-based astrometric observatories can detect or constrain all of these phenomena and improve
upon the expected {\it Gaia} measurements.  Table \ref{Tab:Summary} lists the various proper
motion signals, their VSH modes, the expected amplitude of the 
signal (if known), a recent measurement (if any), and predictions for {\it Gaia}, a long-baseline
next-generation Very Large Array (ngVLA), and a new space astrometry mission.

\hspace{10pt}  
Key observational requirements include all-sky coverage (for low-$\ell$ modes), large extragalactic samples
with reasonable per-object astrometric precision (10--100~$\mu$as), and modest time baselines (5--10 yr).  
These requirements can be met with (1) a ground-based long-baseline radio facility with Very Long Baseline
Array (VLBA) resolution and ten times the collecting area ($100\times 25$~m dishes) or (2) a space-based visible
light astrometric mission with astrometric uncertainties a factor of 3 lower than {\it Gaia} in each
coordinate at any given $G$ magnitude.
The long-baseline radio program would monitor 10,000 objects over 10 years with VLBA-level astrometric
precision: $\pm10$~$\mu$as~yr$^{-1}$ per object.  This sample would be roughly
ten times larger than current VLBA geodetic monitoring samples.  
The optical space-based facility would observe the $\sim10^6$ AGN in the current {\it Gaia} sample.
If the astrometry signal-to-noise could be improved by a factor of 3 compared to
{\it Gaia}\footnote{This could be achieved through a combination of a slightly larger telescope, more
  efficient optics, and a longer mission lifetime.  Proper motion uncertainties scale with time as $t^{-3/2}$.  } 
then the median extragalactic per-source proper motion uncertainty will be 
$\pm100$~$\mu$as~yr$^{-1}$.

\hspace{10pt}  
Either of the above programs would enable detection of global correlated signals of $\sim$0.1~$\mu$as~yr$^{-1}$,
which is $\sim$0.7\% of $H_0$, and we would expect to detect most of the phenomena described in
Section~\ref{sec:signals}.  Gravitational waves, anisotropy, and rotation would likely be (new) upper limits.  
A ground-based radio program may not perform as well as a space-based facility
for measurements requiring fine angular sampling (e.g., BAO evolution) or dense volumetric sampling
(the secular parallax).  This is due to the overall physical paucity of compact radio sources compared to
optical sources across the sky and in the very nearby universe.

\hspace{10pt}  
New ground- or space-based astrometric observatories will provide 
exquisite precision on the Solar motion in the Galaxy and place the strongest constraints to date
on the isotropy of the Hubble expansion in the epoch of dark energy and on the primordial gravitational
wave background over roughly 10 decades in frequency.  We will detect transverse peculiar motion,
extragalactic parallax, and real-time recession of the BAO.

\end{document}